\documentclass[twocolumn,prc,showpacs,floatfix,superscriptaddress,nofootinbib]{revtex4}
\usepackage{graphicx}
\usepackage{pstricks}
\usepackage{dcolumn,amsmath,amssymb}

\begin{document}
\title{Electron-capture branch of $^{100}$Tc and tests of nuclear wave functions for double-$\beta$ decays}
\author{S.~K.~L. Sjue}
\affiliation{Physics Department, 
         University of Washington,
         Seattle, WA 98195, USA}
\author{D. Melconian}
\affiliation{Physics Department,
          Texas A\&M University,
          College Station, TX 77843, USA}
\author{A. Garc\'ia}
\affiliation{Physics Department, 
         University of Washington,
         Seattle, WA 98195, USA}
\author{I. Ahmad}
\affiliation{Argonne National Laboratory, 
             Argonne, 
             Illinois 60439, USA}
\author{A. Algora}
\affiliation{Instituto de Fisica Corpuscular, 
             University of Valencia, 
             Valencia, Spain}
\affiliation{Institute of Nuclear Research, 
             Hungarian Academy of Sciences,
             Debrecen, Hungary}
\author{J. \"Ayst\"o}
\affiliation{Department of Physics, 
             University of Jyv\"askyl\"a, 
             Jyv\"askyl\"a, Finland}
\author{V.-V. Elomaa}
\affiliation{Department of Physics, 
             University of Jyv\"askyl\"a, 
             Jyv\"askyl\"a, Finland}
\author{T. Eronen}
\affiliation{Department of Physics, 
             University of Jyv\"askyl\"a, 
             Jyv\"askyl\"a, Finland}
\author{J. Hakala}
\affiliation{Department of Physics, 
             University of Jyv\"askyl\"a, 
             Jyv\"askyl\"a, Finland}
\author{S. Hoedl}
\affiliation{Physics Department, 
         University of Washington,
         Seattle, WA 98195, USA}
\author{A. Kankainen}
\affiliation{Department of Physics, 
             University of Jyv\"askyl\"a, 
             Jyv\"askyl\"a, Finland}
\author{T. Kessler}
\affiliation{Department of Physics, 
             University of Jyv\"askyl\"a, 
             Jyv\"askyl\"a, Finland}
\author{I.~D. Moore}
\affiliation{Department of Physics, 
             University of Jyv\"askyl\"a, 
             Jyv\"askyl\"a, Finland}
\author{F. Naab}
\affiliation{Nuclear Engineering and Radiological Sciences,
         University of Michigan,
         Ann Arbor, MI 48109, USA}
\author{H. Penttil\"a}
\affiliation{Department of Physics, 
             University of Jyv\"askyl\"a, 
             Jyv\"askyl\"a, Finland}
\author{S. Rahaman}
\affiliation{Department of Physics, 
             University of Jyv\"askyl\"a, 
             Jyv\"askyl\"a, Finland} 
\author{A. Saastamoinen}
\affiliation{Department of Physics, 
             University of Jyv\"askyl\"a, 
             Jyv\"askyl\"a, Finland}
\author{H.~E. Swanson}
\affiliation{Physics Department, 
         University of Washington,
         Seattle, WA 98195, USA}
\author{C. Weber}
\affiliation{Department of Physics, 
             University of Jyv\"askyl\"a, 
             Jyv\"askyl\"a, Finland}
\author{S. Triambak} 
\affiliation{Physics Department, 
         University of Washington,
         Seattle, WA 98195, USA}
\author{K. Deryckx}
\affiliation{Physics Department, 
         University of Washington,
         Seattle, WA 98195, USA}
\date{\today}

\begin{abstract}
We present a measurement of the electron-capture branch of $^{100}$Tc. Our value, $B(\text{EC}) = (2.6 \pm 0.4) \times 10^{-5}$, implies that the $^{100}$Mo neutrino absorption cross section to the ground state of $^{100}$Tc is roughly one third larger than previously thought. Compared to previous measurements, our value of $B(\text{EC})$ prevents a smaller disagreement with QRPA calculations relevant to double-$\beta$ decay matrix elements.
\end{abstract}
\maketitle

\section{Motivation}

If a positive signal were observed from experiments searching for neutrinoless double-beta ($0\nu\beta\beta$) decay, the $\nu$ would be identified as its own anti-particle. In order to extract useful information beyond this important identification, a reliable description of the nuclear wave functions will be essential. For this reason much work has gone into improving the accuracy of nuclear matrix element calculations for double-beta decay \cite{ha:84,ca:96,el:02,ro:03,ci:05}. It is important to test theoretical models by requiring them to reproduce multiple observables that could be sensitive to similar operators. A few double-beta decay candidates, including $^{100}$Mo, have the ground state of the intermediate nucleus with $J^{\pi}=1^+$.  These nuclei allow measurements of single-beta decay rates in addition to the two neutrino double-beta ($2\nu\beta\beta$) decay rates to check calculations.

\begin{figure}
\includegraphics[width=8.0cm,height=6.4cm]{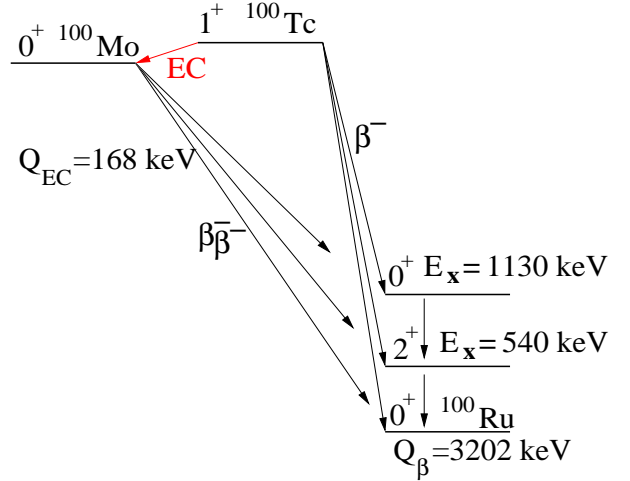}
\caption{The $A=100$ system with its seven experimental observables shown: three $\beta^-$ decays from $^{100}$Tc to $^{100}$Ru, three double-$\beta$ decays from $^{100}$Mo to $^{100}$Ru, and the electron-capture decay from $^{100}$Tc to $^{100}$Mo.}
\label{fig:7obs}
\end{figure}

$^{100}$Mo offers a test system with up to seven constraints (Figure~\ref{fig:7obs}), including measurements of the $2\nu\beta\beta$ decay rate to both the ground state and two excited states of $^{100}$Ru, single-beta decay rates from the intermediate $^{100}$Tc $J^{\pi}=1^+$ state to both the ground state and two excited states of $^{100}$Ru, and the electron-capture (EC) rate from $^{100}$Tc to $^{100}$Mo.  Excluding the highly-suppressed $2\nu\beta\beta$ decay to the $^{100}$Ru $2^+$ excited state, the EC rate is the most uncertain, so a more accurate measurement provides an improved test for theoretical models.

Ejiri {\em et al.}~\cite{ej:00} proposed to use $^{100}$Mo as a detector for both $0\nu\beta\beta$ decay and solar neutrinos. For the latter, the efficiency for low-energy neutrino captures is determined by the same matrix element that drives the rate for the EC transition from $^{100}$Tc to $^{100}$Mo. The basic features of the detector can be found in Ref.~\cite{ej:00}. Here we address only the effect of our measurement on the amount of Mo necessary to make it a sufficiently efficient detector of solar neutrinos. Ref.~\cite{ej:00} concluded that the amount of $^{100}$Mo needed to perform a significant measurement would be $3.0\times 10^3~$kg of $^{100}$Mo ($31\times 10^3~$kg of natural Mo).  Their calculation was based on an indirect determination of the strength for the transition:
\begin{equation}
B({\rm GT};^{100}{\rm Mo} \rightarrow ^{100}{\rm Tc}) = 3 g_A^2
|\langle ^{100}{\rm Tc} || \sigma \tau || ^{100}{\rm Mo} \rangle |^2
\end{equation}
deduced from a $^3{\rm He} + ^{100}{\rm Mo} \rightarrow ^3{\rm H} + ^{100}{\rm Tc}$ measurement~\cite{ak:97} which yielded:
\begin{equation}
B(\text{GT})_{\text{indirect}} = 0.52 \pm 0.06.
\end{equation}
However, determinations of the weak strength via charge-exchange reactions can be inaccurate~\cite{ga:95}. It is possible to directly determine the branching ratio.  A previous experiment~\cite{ga:93} measured the $^{100}$Tc EC branch to be $(1.8 \pm 0.9) \times 10^{-5}$ from which one obtains $B(\text{GT};^{100}\text{Mo} \rightarrow ^{100}\text{Tc}) = 0.66 \pm 0.33$, barely inconsistent with zero. Here we present a more precise measurement of the EC branch and discuss its implications. 

\section{Apparatus}

The experiment was performed using the IGISOL~\cite{jr:01,jh:04} facility at the University of Jyv\"askyl\"a. A proton beam delivered from the K130 Cyclotron with $E_p = 10$ MeV and intensity $I \sim 24~\mu$A impinged on a $\rho \approx500~\mu$g/cm$^2$, 97.4\%-enriched $^{100}$Mo target which was placed in an ion guide with helium at $p \approx100~$mbar. The $^{100}$Tc ions recoiled into the helium where they thermalized and the fraction that remained ionized were subsequently extracted from the gas cell.  All ions were electrostatically guided through an RF sextupole ion beam guide while the neutral gas was differentially pumped away.  Finally, the ions were accelerated toward the mass separator at an electrostatic potential of $\phi \approx30~$kV. 

\begin{figure}
\includegraphics[width=8.0cm,height=4.9cm]{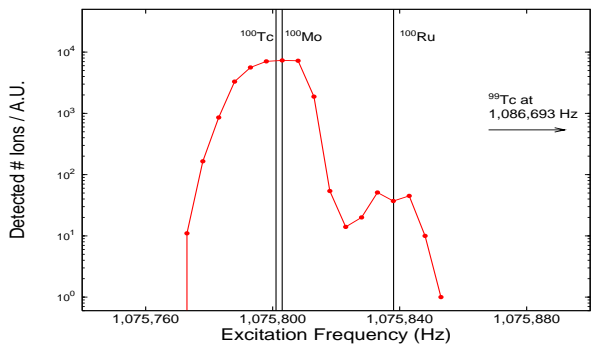}
\caption{Mass scan to show the mass resolving power for $A=100$ obtained with JYFLTRAP.   }
\label{fig:mass100}
\end{figure}

The $A=100$ component of this beam was roughly separated by a magnet with a mass-resolving power of $M/\Delta M \approx250$, after which it was cooled and bunched in a linear segmented RFQ trap~\cite{an:02}.  The bunched beam was introduced into a Penning trap in a $7-$T magnetic field with a helium buffer gas~\cite{vk:04}, in which isobaric purity was achieved by means of a mass-selective buffer gas cooling technique~\cite{gs:91}.  Figure~\ref{fig:mass100} shows a mass scan for $A=100$ from the purification trap with mass resolving power of $M/\Delta M \approx 25,000$, more than enough to prevent contamination from $^{99}$Tc, which comes with unwanted Tc x rays from an isomeric state with a long half-life, $t_{1/2} \approx 6~$h.  The excitation frequency was set to $f = 1,075,800~$Hz for beam purification during the experiment.  

The purified $A=100$ beam was extracted from the trap and implanted inside a scintillator designed to achieve $>99$\% coverage while allowing the implantation spot to be as close as $\approx0.32~$cm to a Ge detector.  Decay rates as high as $R \approx20~$kHz were observed by the scintillator during the course of the experiment.  A hollowed cylinder within the scintillator held vacuum as part of the same volume as the ion trap.  Ions from the trap stopped in a $\approx25~\mu$m-thick aluminum foil that was inserted into the scintillator.  A $\approx6~$mm-diameter collimator mounted on the foil holder prevented deposition of ions onto the sidewalls of the cylinder inside the scintillator.  This was checked by implanting $^{99}$Tc$^m$ into the foil, cutting the foil into pieces, then monitoring the end and sides of the foil for the 140-keV $\gamma$ rays from the $t_{1/2}\approx 6~$h isomeric state.  Activity was only found on the end of the foil.

\begin{figure}
 \includegraphics[width=8.0cm,height=4.9cm]{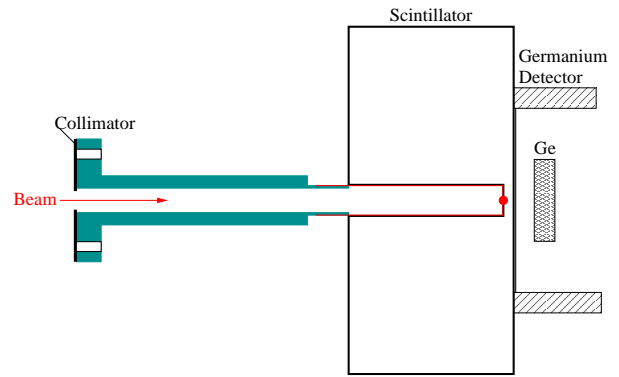}
  \caption{Experimental setup: $^{100}$Tc ions were deposited onto an aluminum foil inside the scintillator. The beam was tuned by triggering on the scintillator.  The scintillator allowed veto of $>90$\% of $\beta$-decay events.  The foil in which the ions stopped was only separated from the Ge detector by the $\approx0.32~$cm-thick face of the scintillator to maximize photon-detection efficiency.  The dot inside the scintillator near the Ge detector represents activity deposited in the aluminum foil.}
  \label{fig:setup}
\end{figure}  

A sketch of the counting setup is shown in Figure~\ref{fig:setup}.  A $10~$mm-thick, $25~$mm-diameter Ge (LEPS) detector abutted the scintillator. The Ge detector had an energy resolution of FWHM$\approx420~$eV at $E_\gamma \approx17~$keV and solid-angle coverage of $\approx~17\%$ of $4\pi$. The scintillator detector, which produced signals from $\beta$-particles emitted in the decay to $^{100}$Ru, enabled efficient veto of backgrounds from low-energy $\beta$s and Ru x rays in the Ge detector.  Signals from the scintillator were read with two PMTs optically coupled to opposing faces of the scintillator perpendicular to the beam axis.

We produced two amplifications of the Ge detector signal: one with high gain, to observe the x rays with sufficient resolution, and one with low gain to measure $\gamma$ rays.  With every event, we recorded these two signals, the amplitudes of the signals from two phototubes on the scintillator, and TAC signals between x rays and either phototube.  Any signal with amplitude larger than $2.4~$keV in the x-ray detector triggered data acquisition.  One signal from every 999 scintillator signals also triggered data acquisition, to allow an independent measurement of the number of decays.

\section{Results and Analysis}

The EC branch is determined by the ratio of the number of EC decays, $N_{\text{EC}}$, to the total number of decays, $N_{tot}$.  Mo K-shell x rays signal EC events from which we calculate $N_{\text{EC}}$.  Measurements of the $539.6$- and $590.8$-keV $\gamma$-ray intensities, together with a calibration of the Ge detector's efficiency, allow us to determine $N_{tot}$ from the number of counts in either the $539.6$- or the $590.8$-keV $\gamma$-ray lines.   

\subsection{Photon Efficiency}
\label{sec:efficiency}

\begin{figure}
\includegraphics[width=8.0cm,height=6.0cm]{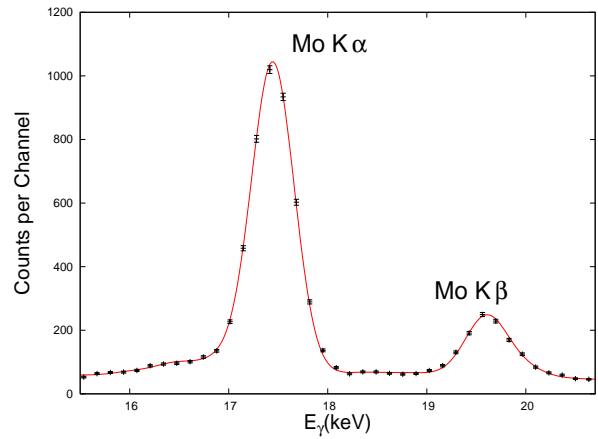}
\caption{$^{92}$Tc x-ray spectrum. The line is a fit to one overall amplitude, with the relative areas of the peak determined by the product of the known intensities~\cite{bf:96} and efficiencies determined by simulations using PENELOPE~\cite{penelope}.  The simulations were also used to fix the shape of the Compton shoulder, which is visible below the K$\alpha$ x ray.  These data were used to get the relative efficiency between Mo-K$\alpha$ and Mo-K$\beta$ x rays.}
\label{fig:Tc92}
\end{figure}

The relative efficiency between the Mo K-shell x rays and the $539.6$-keV and $590.8$-keV $\gamma$ rays,  needed to extract $N_{\text{EC}}/N_{\text{tot}}$, was obtained from calibration sources and simulations based on the experimental geometry.  $^{92}$Tc was obtained from $^{92}$Mo impurities in the enriched $^{100}$Mo target by tuning the dipole magnet, RFQ buncher, and Penning trap for $A=92$. Figure~\ref{fig:Tc92} shows the spectrum of x rays with our fit for the $A=92$ beam. 

We produced fits using a line-shape functional consisting of a low-energy exponential folded with a Gaussian, plus a low-energy shoulder for Compton-scattered x rays with a shape determined by PENELOPE~\cite{penelope} simulations.  In our fits we fixed the relative x-ray intensities and extracted the relative efficiencies for the two Mo-K$\alpha$ and three Mo-K$\beta$ x rays.  The shape of the Compton shoulder and calibration used in Figure~\ref{fig:Tc92} were also used to constrain all fits to the x-ray spectra from $^{134}$Cs and $^{100}$Tc.  The relative efficiency between the Mo-K$\alpha$ and K$\beta$ x rays is dominated by the dead layer from the contact at the front of the Ge crystal and the thickness of the thin wall of the scintillator (see Figure~\ref{fig:setup}).  

\begin{figure}
\includegraphics[width=8.0cm,height=5.8cm]{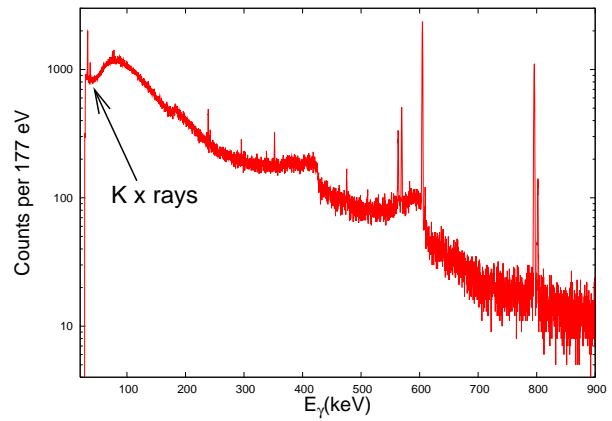}
\caption{$^{134}$Cs $\gamma$-ray spectrum to get the relative efficiency between x rays and the $539.6$-keV and $590.8$-keV transitions.}
\label{fig:effi}
\end{figure}

Figure~\ref{fig:effi} shows a $^{134}$Cs-source $\gamma$-ray spectrum from a calibration source made by evaporating a solution with $^{134}$Cs on a foil made to fit in our scintillator.  We used the known $^{134}$Cs x-ray and $\gamma$-ray intensities~\cite{as:04} to determine the relative efficiencies between x and $\gamma$ rays.

\begin{figure}
\includegraphics[width=8.0cm,height=5.8cm]{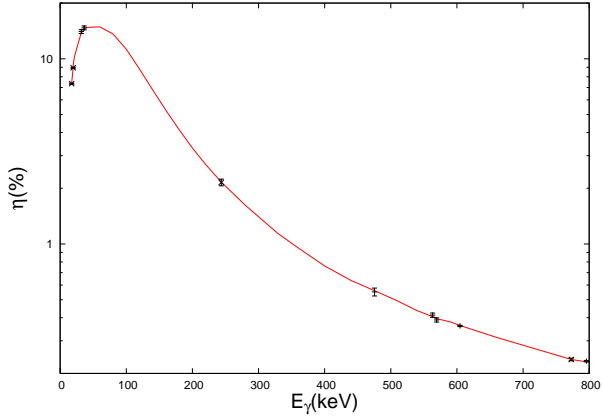}
\caption{Efficiencies determined using x rays and $\gamma$ rays from the $^{92}$Tc (points with x at center), x rays and $\gamma$ rays from $^{134}$Cs (points with - at center), and Monte Carlo simulations using the code PENELOPE~\cite{penelope} (red line).  The two points at 17.44 keV and 19.65 keV from $^{92}$Tc are identical to the x rays that signal the EC decay of $^{100}$Tc.}
\label{fig:effit}
\end{figure}

Figure~\ref{fig:effit} shows the efficiencies for x rays and $\gamma$ rays from both $^{92}$Tc and $^{134}$Cs with the results of Monte Carlo simulations performed using the code PENELOPE~\cite{penelope}.  The 563.2- and 569.3-keV $\gamma$ rays from $^{134}$Cs are conveniently close in energy to the 539.6- and 590.8-keV $\gamma$ rays from the decay of $^{100}$Tc.  The simulations were used to perform the interpolation necessary to determine the relative efficiencies between the $^{100}$Ru $\gamma$ rays and the Mo K-shell x rays used in our branch calculation.  

\subsection{Electron-Capture Branch Calculation}

\begin{figure}
\includegraphics[width=8.0cm,height=5.8cm]{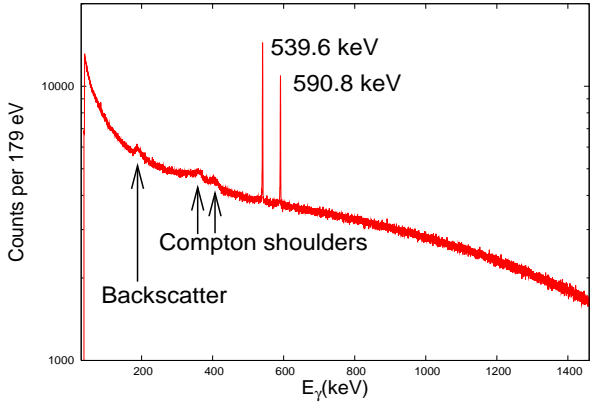}
\caption{$\gamma$-ray spectrum from $^{100}$Tc beam, showing only a Pb x ray from lead shielding at 74 keV, the 539.6- and 590.8-keV $\gamma$ rays and their Comptons, and a continuous $\beta$ background.}
\label{fig:gamma_spectrum}
\end{figure}  

\begin{figure}
\includegraphics[width=8.0cm,height=5.8cm]{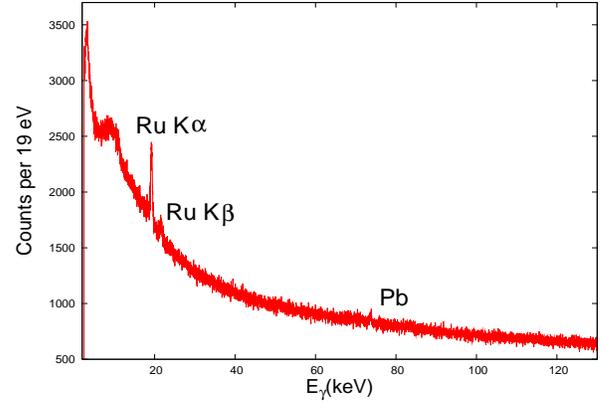}
\caption{Raw x-ray spectrum from $^{100}$Tc beam.  The Ru-K$\alpha$ and Ru-K$\beta$ lines are visible at 19.2 keV and 21.6 keV.  A Pb x ray from lead shielding is visible at 74 keV.}
\label{fig:xray_spectrum}
\end{figure}

\begin{figure}
\includegraphics[width=8.0cm,height=5.8cm]{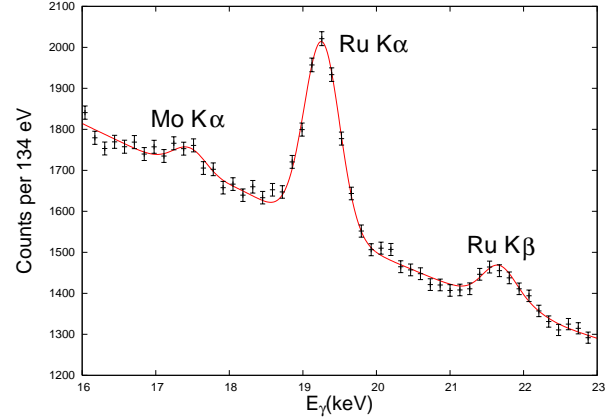}
\caption{Fit to vetoed x-ray spectrum for five runs.  Each x-ray peak from both isotopes was constrained during the fit to have an area equal to the product of one overall amplitude for the isotope, the fluorescence yield for each peak, and the efficiency of the LEPS detector at each peak's energy.  The Mo K$\alpha$ peak is at 17.7 keV, the Ru K$\alpha$ peak is at 19.2 keV, and the Ru K$\beta$ peak is at 21.6 keV.  This fit yields $\chi^2/\nu=1.042$ with $\nu=394$.}
\label{fig:Mo-fit}
\end{figure} 

Figure~\ref{fig:gamma_spectrum} shows a raw $\gamma$-ray spectrum taken with the $^{100}$Tc beam.  Figure~\ref{fig:xray_spectrum} shows a raw x-ray spectrum taken with the $^{100}$Tc beam.  Figure~\ref{fig:Mo-fit} shows the fit for the Mo- and Ru-x-ray lines to a scintillator-vetoed x-ray spectrum from five runs.  We calculate the electron-capture branch as:
\begin{equation}
B(\text{EC}) = \frac{A(\text{Mo-K})}{A(\text{590.8-keV})}
             \frac{\eta(\text{590.8-keV})}{\eta(\text{Mo-K})}
             \frac{(1-c) I_\gamma(\text{590.8-keV})}{f_{K}\omega_{K}},
\label{eq:branch}
\end{equation}
where $A(\text{Mo-K})$ and $A(\text{590.8-keV})$ are the photopeak areas for the Mo-K and 590.8-keV transitions; $\eta(\text{590.8-keV})/\eta(\text{Mo-K})$ is the relative efficiency between the 590.8-keV and Mo-K transitions; $c$ is the fraction of 590.8-keV $\gamma$ rays lost because of summing from coincident $\beta$-particles and 539.6-keV $\gamma$ rays, calculated from the same simulations used to determine the Ge detector efficiency; $I_\gamma(\text{590.8-keV})$ is the absolute intensity of the 590.8-keV $\gamma$ ray; $f_K = 0.88$ is the fraction of EC decays that produce a vacancy in the K shell; and $\omega_{K} = 0.765$ is the total K fluorescence yield~\cite{bf:96}, i.e., the probability of emission of a K-shell Mo x ray per K-shell vacancy.  In practice, because the efficiency changes between the K$\alpha$ and K$\beta$ lines, we obtained $A(\text{Mo-K})/\eta(\text{Mo-K})$ as the sum of $A(\text{Mo-K}_i)/\eta(\text{Mo-K}_i)$ over all the individiual K-shell lines.  

Over the course of several days running the experiment, we observed changes in the Ge detector's resolution.  In our analysis we independently determined the branch from every run with a resolution better than FWHM$\leq700$ eV in the x-ray region.  

\begin{figure}
\includegraphics[width=8.0cm,height=5.8cm]{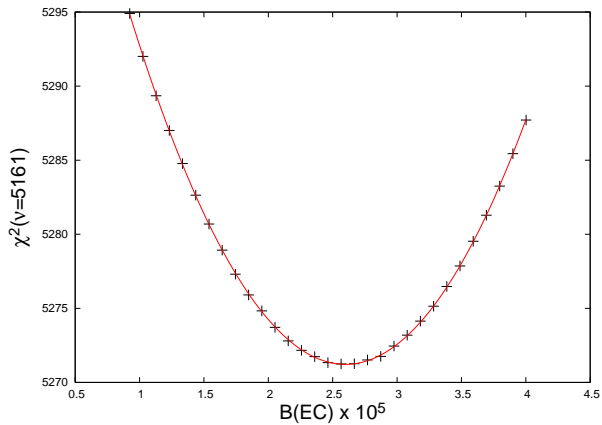}
\caption{Plot of the cumulative $\chi^2$ from 13 runs as a function of the assumed $B(\text{EC})$.}
\label{fig:brfit}
\end{figure}

To get the best value of the $B(\text{EC})$ from all runs, including short runs from which one would individually obtain a value of $B(\text{EC})$ statistically consistent with zero, we used the following scheme.  For an assumed $B(\text{EC})$, we calculated the number of Mo x rays expected given the number of 590.8-keV $\gamma$ rays, fit the vetoed x-ray spectra from all runs with the Mo x-ray areas fixed, and took the total $\chi^2$ from all runs.  Figure~\ref{fig:brfit} shows a plot of the results, from which we obtain
\begin{equation}
B(\text{EC})=(2.60 \pm 0.34 \pm 0.20)\times 10^{-5},
\end{equation}
in which the first uncertainty is statistical and the second uncertainty is due to the Ge detector calibration.  For the analysis that follows, we combine the uncertainties and use the result $B(\text{EC})=(2.6 \pm 0.4)\times10^{-5}$.

This result is more precise than the previous determination~\cite{ga:93}: $B(\text{EC}) = (1.8 \pm 0.9) \times 10^{-5}$.  That experiment did not use a high-resolution mass separator and consequently had to make a separate measurement to determine the contributions from contaminants. Radioactivity was collected on a tape for several hours, then $\gamma$ rays from the unwanted isotopes were measured, and finally the number of Mo K-shell x rays due to the electron capture of $^{100}$Tc during the experiment was deduced by accounting for veto efficiencies, branching ratios, and the effect of the periodic movement of the tape. 

\subsection{Calculation of $\gamma$-ray Intensities}

\begin{figure}
\includegraphics[width=8.0cm,height=5.8cm]{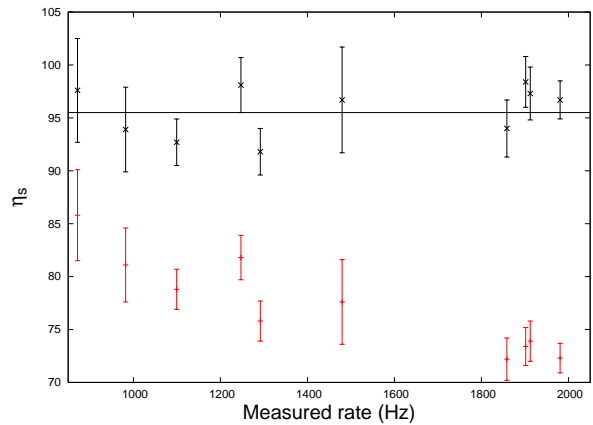}
\caption{Scintillator efficiency as a function of data acquisition rate.  The red points are prior to correction, the black points are corrected, and the overall fit of $\eta_s(E_x=1130~\text{keV})=95.5 \pm 0.8$\% gives $\chi^2/9=0.895$.  The actual rates that determined the dead time correction were the scintillator rates, which were approximately an order of magnitude larger, approaching 20 kHz at its maximum.}
\label{fig:scinteff}
\end{figure}

We use the 590.8-keV $\gamma$ ray to determine the efficiency of the scintillator.  Direct $\beta$-decay feeding of the $^{100}$Ru excited state with $E_x=1130~$keV accounts for 99.8\% of the 590.8-keV $\gamma$-ray intensity~\cite{100elevels}, which makes it convenient to determine the scintillator's efficiency for a known $\beta$-decay energy spectrum.  To do this, we gate on the photopeak of the 590.8-keV $\gamma$ ray and find the number of TAC signals.  Each $\gamma$ ray is the result of a $\beta$ decay, so the efficiency of the scintillator for this decay branch should be 
\begin{equation}
\eta_s(E_x=1130~\text{keV}) = \frac{A(\text{TAC})}{A(\gamma)},
\end{equation}
where $A(\text{TAC})$ is the number of coincidences between the Ge detector and the scintillator.  Because we used a 19-$\mu$s delay on scintillator signals that stopped the TAC to make them come after the slower signals from the Ge preamplifier, $A(\text{TAC})$ required a dead-time correction.  Ten periods from all the data were selected for approximately constant activity rates, then corrected.  Figure~\ref{fig:scinteff} shows the resulting measurement of the scintillator's efficiency, which gives the result $\eta_s(E_x=1130~\text{keV}) = 95.5 \pm 0.8$\%.  

The simulation geometry used to determine our relative photon efficiency, $\eta(E_\gamma)$, was also used to calculate the efficiency of the scintillator for this $\beta$-decay branch, yielding $\eta_{\text{sim}}(E_x=1130~\text{keV})=95.4\%$.  This simulation was performed without tuning any of the simulation parameters, using the design specifications for the scintillator's dimensions and the measured thickness of the aluminum target foil.  Another simulation for the $\beta$ decay directly to the ground state of $^{100}$Ru gives $\eta_{\text{sim}}(E_x=0~\text{keV})=98.1\%$.  Simulations of all the $\beta$-decay branches, using the intensities given in Ref.~\cite{100nds}, yield an average efficiency for all $^{100}$Tc $\beta$ decays of ${\bar \eta}_{\text{sim}}=97.9\%$.  While this value depends on the assumed branches, the dependence is very small, as indicated by the difference of only 0.2\% compared to the ground state branch.  Propagating the uncertainty in the measured efficiency for $\beta$ decay to the $E_x=1130~$keV excited state to the average efficiency calculated from simulations, we obtain ${\bar \eta}=97.9\pm0.8$. 

Given the scintillator's efficiency, the number of triggers from the scintillator $N_{st}$, and recalling that the scintillator triggers were divided by 999, we obtain the number of decays for the same rate-selected data as 
\begin{equation}
N_D = \frac{999 \cdot N_{st}}{\bar \eta}.
\end{equation}
To calculate the absolute intensities of the $\gamma$ rays, we use
\begin{equation}
I_\gamma = \frac{A(\gamma)}{N_D \eta(E_\gamma) (1-c)},
\label{eq:Igamma}
\end{equation}
from which we obtain
\begin{equation}
I_\gamma(\text{590.8-keV})=5.5\pm0.3\%
\end{equation}
and
\begin{equation}
I_\gamma(\text{539.6-keV})= 6.6\pm0.3\%.
\end{equation}
The uncertainties are due to uncertainties in the Ge detector calibration.  These intensities can be compared with the recent precision measurements~\cite{100gbrs1,100gbrs2} of $I_{\gamma}(\text{590.8-keV})=5.5\pm0.3\%$ and $I_\gamma(\text{539.6-keV})=6.6\pm0.5\%$.  The agreement is remarkable, considering the estimated 4.5\% uncertainty in our calibration of the Ge detector's absolute efficiency.  

\subsection{Internal Ionization and Excitation}

Our experiment also allows us to determine the probability of K-shell internal ionization and excitation (IIE)~\cite{Isozumi} from the decay of $^{100}$Tc.  The Ru K$\alpha$ x-ray peak in the raw x-ray spectrum originates mainly from three sources: internal conversion (IC) of the 539.6- and 590.8-keV $\gamma$ rays, and IIE from the $\beta^-$ decay of $^{100}$Tc.  The tabulated IC coefficients from Ref.~\cite{bf:96}, the probability of a K-shell vacancy due to IC per $\gamma$ ray emitted, are $e_K/\gamma(539.6)=(3.8\pm0.2)\times 10^{-3}$ and $e_K/\gamma(590.8)=(3.0\pm0.2)\times 10^{-3}$.  Our data and calibration allow us to determine the probability of K-shell vacancy due to IIE per $\beta^-$ decay, $P_K$, from the number of Ru K-shell x rays versus the number of $\gamma$ rays.  We find
\begin{equation}
P_K=(7.2\pm0.6)\times 10^{-4},
\end{equation} 
which can be compared with a measurement of $P_K=(6.0\pm0.6)\times 10^{-4}$ from Ref.~\cite{ga:93}.  Note that Ref.~\cite{ga:93} used different IC coefficients; the results would show better agreement if the same IC coefficients were used.

\subsection{Systematic Uncertainties}

The purification Penning trap ensured that only ions having $A=100$ could reach the experimental setup.  Neither the $\gamma$-ray nor x-ray spectra (Figure~\ref{fig:gamma_spectrum} and Figure~\ref{fig:xray_spectrum}) show any signs of contaminants.  

Mo x rays could potentially be generated by fluorescence of $^{100}$Mo, coming with the $A=100$ beam and from the decay of $^{100}$Tc. We inserted a 1 $\mu$m-thick Pd foil between the scintillator and Ge detector to check for fluorescence  while taking the $A=100$ beam.  The amount of Pd in the foil is $\approx10^{10}$ times greater than the total amount of Mo deposited during the entire experiment.  No Pd x rays were observed; thus we exclude contamination of the Mo x rays by fluorescence.  

Our calibration scheme determines the Ge detector efficiency as a function of photon energy.  We used the same simulations used for the efficiencies to determine the summing corrections used in our determination of the $\gamma$-ray intensities ($c$ in Eq.~\ref{eq:Igamma}).  Uncertainties in the actual geometry of the experiment, including detector specifications for both the scintillator and Ge detector, could cause these values to be inaccurate.  To account for these geometrical uncertainties, we calculated an uncertainty based on a shift in the detector's beam-axis position of 0.5 mm for both the summing corrections and the Ge efficiency, $\eta(E_\gamma)$.  We also studied uncertainties due to radial beam position and beam spread, which we found to be negligible.

For the calculation of $B(\text{EC})$ (Eq.~\ref{eq:branch}), the uncertainty in the summing correction $c$ and the $\gamma$-ray efficiency $\eta(E_\gamma)$ are negligible because coincidence measurements with the scintillator determine $\eta(\text{590.8-keV})(1-c)I_\gamma(\text{590.8-keV})$ to 1\% accuracy.  The uncertainty in $\eta(\text{Mo-K})$ was determined from the fits explained in Sec~\ref{sec:efficiency} to be 6.2\%.  This was added in quadrature to smaller effects due to experimental geometry and beam variations described above to determine an overall systematic uncertainty of 7.7\% in our determination of $B(\text{EC})$.  The same systematic uncertainty applies to the determination of $P_K$ for IIE.

For our determination of $I_\gamma(\text{590.8-keV})$ and $I_\gamma(\text{539.6-keV})$, the estimated error due to $\eta(E_\gamma)$ is 4.5\%.  The efficiencies for the higher energy x rays in the $^{134}$Cs spectrum (Fig.~\ref{fig:effi}), which were used to determine the ratios between the efficiencies for the x rays and $\gamma$ rays, show less sensitivity to the parameters tuned in our simulations.  Corrections calculated for both the 539.6- and 590.8-keV $\gamma$ rays included summing from both $\beta$-particles and the angular correlation between the $E2$ transitions in the $0^+\rightarrow2^+\rightarrow0^+$ $\gamma$-ray cascade.  
\section{Conclusions}

Our determination of $B(\text{EC})$ implies $\log{ft}=4.29^{+0.08}_{-0.07}$ for the EC decay of $^{100}$Tc.  We used the measurement $t_{1/2}=15.27\pm0.05~$s from Ref.~\cite{100gbrs1}, the most precise determination of the $^{100}$Tc lifetime available, along with $f=0.331(24)$ calculated from the tables in Ref.~\cite{bf:96}, to determine this value. This can be compared with the $\log{ft}$s for the decays of $^{98}$Zr and $^{102}$Mo, for which the $\log{ft} \approx4.2$.  

Our determination of the EC branch gives the Gamow-Teller strength,
\begin{equation}
B(\text{GT}; ^{100}\text{Mo} \rightarrow ^{100}\text{Tc})= 0.95 \pm 0.16,
\end{equation}
which is approximately 80\% larger than the value $B(\text{GT};^{100}\text{Mo} \rightarrow ^{100}\text{Tc})=0.52\pm0.06$ estimated using the charge exchange reaction~\cite{ak:97}.  Revising the estimate of Ref.~\cite{ej:00} based on our measurement, a solar neutrino detector would require $1.6\times 10^3~$kg of $^{100}$Mo ($17\times10^3~$kg of natural Mo).

With respect to testing calculations of nuclear matrix elements for double-beta decays: QRPA predictions~\cite{gr:92,su:94} for the transition strength are in the range $4 \le B(\text{GT}; ^{100}\text{Mo} \rightarrow ^{100}\text{Tc}) \le 6$, so disagreement remains significant.  A recent paper by Faessler \textit{et al}~\cite{fa:08} shows that the full set of observables may be reproduced by fitting the axial vector coupling constant $g_A$ and allowing values smaller than 1.

If we assume that the $2\nu\beta\beta$-decay rates are dominated by the ground state contribution, we can determine the $2\nu\beta\beta$ half lives from the equation 
\begin{equation}
T_{1/2}^{-1}=f |M_{2\nu}|^2,
\end{equation} 
in which the $M_{2\nu}$ is determined by
\begin{widetext}
\begin{equation}
M_{2\nu}=\frac{\langle^{100}\text{Ru}||\tau\sigma||^{100}\text{Tc}(\text{gs})\rangle\langle^{100}\text{Tc}(\text{gs}) || \tau\sigma ||^{100}\text{Mo}\rangle}{M_{\text{Mo}}-E_{\beta 1}-E_{{\bar \nu}_{e1}}-M_{\text{Tc}}}.
\end{equation} 
\end{widetext}
The simplest approximation is to assume that $\langle E_{\beta 1}+E_{{\bar \nu}_{e1} }\rangle=Q_{\beta\beta}/2$.  The phase space integrals were performed without this approximation in Ref.~\cite{SSD100Mo}.  Table~\ref{tab:2nu} reproduces the calculated values of  $T_{1/2}^{2\nu\beta\beta}$ from our measurement of $B(\text{EC})$ for the approximate denominator and using the calculations of Ref.~\cite{SSD100Mo}, compared with available measurements.  The ground state alone predicts a larger $2\nu\beta\beta$ decay rate than the actual measurement for both measured decays to $0^+$states in $^{100}$Ru.  This shows that the ground state plays an important role in the $2\nu\beta\beta$ decay rates.
\begin{table}
\begin{tabular}{llccc}
\hline\hline
\multicolumn{2}{c}{$^{100}$Ru level}& $T_{1/2}^{2\nu\beta\beta}$-SSD1 & $T_{1/2}^{2\nu\beta\beta}$-SSD2 & $T_{1/2}^{2\nu\beta\beta}$-Exp \\
$J^\pi$&$E_x(\text{keV})$& (years) & (years) & (years) \\
\hline
$0^+$& 0 & $6.2(9)\times 10^{18}$&$5.0(7)\times 10^{18}$ &$7.3(4)\times 10^{18}$ \\
$0^+$&1130 &$3.8(6)\times 10^{20}$&$3.1(5)\times 10^{20} $&$5.7(^{+1.5}_{-1.2})\times 10^{20}$ \\
$2^+$&539.6&$3.2(5)\times 10^{23}$&$1.2(2)\times 10^{23}$&$>1.1\times10^{21}$ \\
\hline\hline
\end{tabular}
\caption{Predictions of the single-state dominance hypothesis versus experimental data for $2\nu\beta\beta$ decays of $^{100}$Mo.  The first column (SSD1) uses the approximation $\langle E_{\beta_1}+E_{{\bar \nu}_{e1}}\rangle=Q_{\beta\beta}/2$.  The second column (SSD2) includes the integrated denominator from Ref.~\cite{SSD100Mo}.  The third column lists experimental data from Ref.~\cite{pdg08} for comparison.}
\label{tab:2nu}
\end{table}

\acknowledgments

We thank Jerzy Szerypo and Peter Dendooven for help in the initial stages of this experiment.  This work was partly supported by the US Department of Energy, under contracts DE-F602-97ER41020 at the University of Washington and W-31-109-ENG-38 at Argonne.
This work has been supported through a European Community Marie Curie
Fellowship and by the Academy of Finland under the project No. 202256
and the Centre of Excellence programme 2006-2011 (Nuclear and Accelerator Based
 Physics Programme at JYFL).
A. Algora gratefully acknowledges support of the Janos Bolyai and Ramon y Cajal research fellowships, as well as support from EC contract MERG-CT-2004-506849 and MEC-FPA2005-03993.

\end{document}